\documentclass[prl,letterpaper, reprint, amsmath, amssymb,lengthcheck, showpacs, superscriptaddress, twocolumn, floatfix]{revtex4-1}
\usepackage{graphicx}
\usepackage{slashed}
\usepackage{xcolor}

\usepackage{mathtools}
\usepackage{amsmath}
\usepackage{comment} 
\usepackage{textcomp}

\usepackage{bm}

\usepackage{enumitem}

\usepackage[utf8]{inputenc}

\def\bea{\begin{eqnarray}}
\def\eea{\end{eqnarray}}

\def\lmatrix{\left(\begin{array}}
	\def\rmatrix{\end{array}\right)}

\def\msbar{\overline{\rm MS\kern-0.5pt}\kern0.5pt}

\begin{document}
	
	\title{Fate of the conformal fixed point with twelve massless fermions and SU(3) gauge group}
	
	\author{Zoltan Fodor}
	\affiliation{University of Wuppertal, Department of Physics, Wuppertal D-42097, Germany,\\
	Juelich Supercomputing Center, Forschungszentrum Juelich, Juelich D-52425, Germany}
	
	\author{Kieran Holland}

	\affiliation{University of the Pacific, 3601 Pacific Ave, Stockton CA 95211, USA}
	
	\author{Julius Kuti}
	\affiliation{University of California, San Diego, 9500 Gilman Drive, La Jolla, CA 92093, USA}
	
	\author{Santanu Mondal}
	\author{Daniel Nogradi}
	\affiliation{E\"otv\"os University, Institute for Theoretical Physics, 
		MTA-ELTE Lendulet Lattice Gauge Theory Research Group,  Budapest 1117, Hungary}
	
	\author{Chik Him Wong}
	\affiliation{University of Wuppertal, Department of Physics, Wuppertal D-42097, Germany}

\begin{abstract}

We report new results
on the  conformal properties of an important strongly coupled gauge theory, a building 
block of composite Higgs models 
beyond the Standard Model.
With twelve massless fermions in the fundamental representation of the SU(3) color gauge group,
an infrared fixed point of the $\beta$-function was recently reported in the theory~\cite{Cheng:2014jba} 
with uncertainty 
in the location of the critical gauge coupling inside the narrow $[ 6.0<g_*^2<6.4]$  interval
and widely accepted since as the strongest evidence 
for a conformal fixed point and scale invariance in the theory with model-building 
implications.
Using the exact same renormalization scheme as the previous study, we show that 
no fixed point of the $\beta$-function exists in the reported interval.
Our findings eliminate the only seemingly credible evidence for conformal fixed 
point and scale invariance in the $N_f=12$  model whose infrared properties remain unresolved.
The implications of the recently completed 5-loop QCD beta function 
for arbitrary flavor number are discussed with respect to our work.

\end{abstract}	
	
	
\maketitle

{\begin{center}\bf{I. INTRODUCTION AND MOTIVATION}\end{center}} 

Investigations of strongly coupled gauge theories with massless fermions 
in the fundamental or two-index symmetric (sextet) representation 
of the SU(3) color gauge group serve considerable theoretical interest with added relevance as
important building blocks of composite Higgs theories beyond the Standard Model (BSM). Two 
complementary aspects of the composite Higgs paradigm are investigated 
in this large class of theories: (1) a near-conformal and
unexpectedly light scalar particle, 
perhaps dilaton-like with mass at the Electroweak scale 
or (2) a parametrically light pseudo Nambu-Goldstone boson (PNGB) combined
with partial compositeness for fermion mass generation to avoid the flavor problem.
Both paradigms are based on strongly coupled gauge dynamics 
to address important aspects of conformal and chiral symmetries and their symmetry breaking patterns
in BSM theories. 
The precise determination of near-conformal or conformal behavior of 
SU(3) gauge theory with twelve flavors is relevant for both paradigms.

\begin{figure}[h!]
	\vskip -0.15in
	\includegraphics[width=1.1\linewidth]{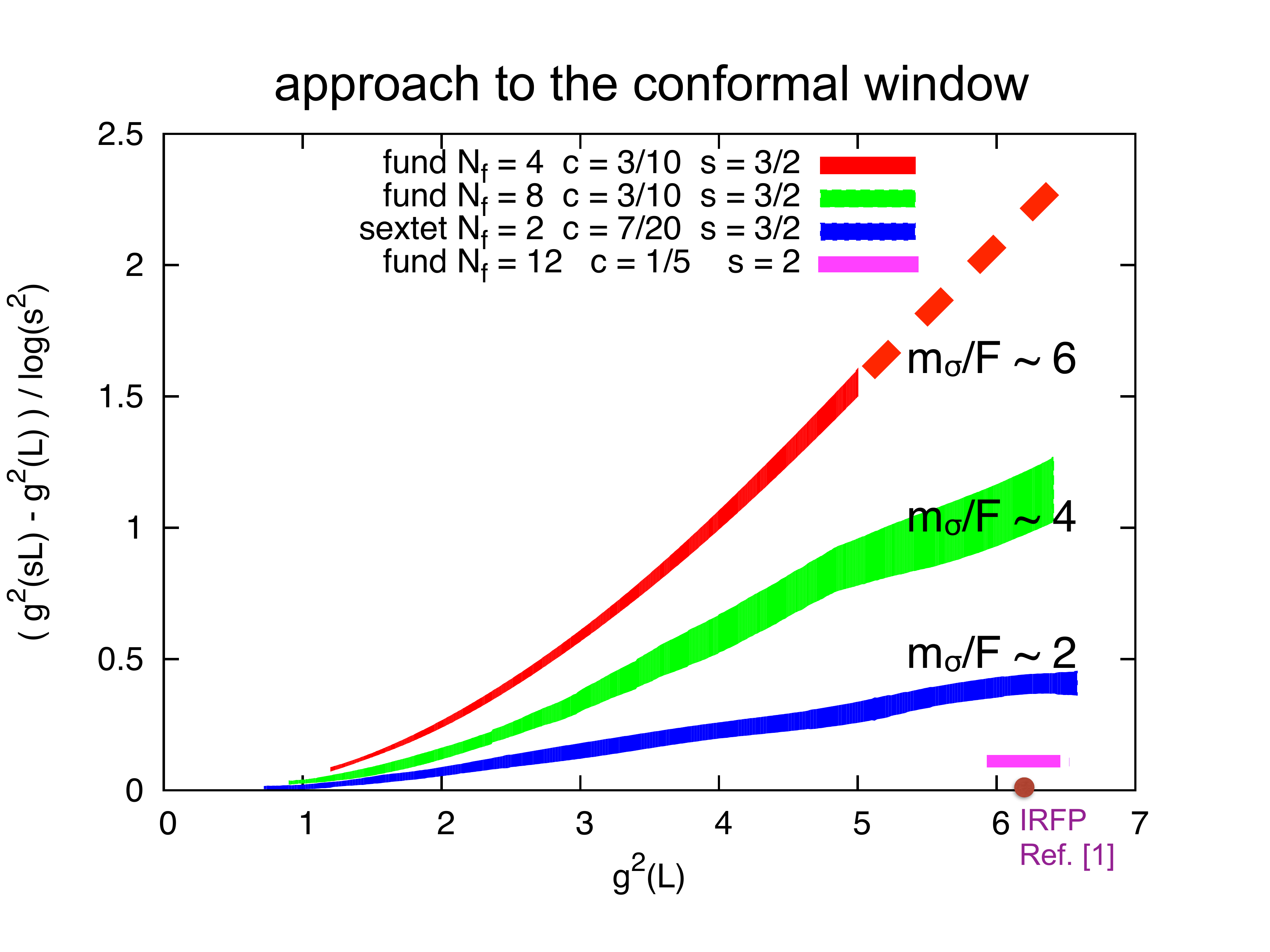}
	\vskip -0.2in
	\caption{The step $\beta$-functions of strongly coupled gauge theories in two 
		different fermion representations of the SU(3) gauge group are color coded. 
		The $N_f=4$ $\beta$-function is from~\cite{Fodor:2012td} (dashed line segment extrapolated)
		with the $m_\sigma/F$ ratio taken from QCD,  the
		$N_f=8$ $\beta$-function is  from~\cite{Fodor:2015baa} with the $m_\sigma/F$ ratio 
		from~\cite{Aoki:2014oha,Appelquist:2016viq}, 
		and the sextet $\beta$-function is from~\cite{Fodor:2015zna} with the $m_\sigma/F$ ratio 
		taken from~\cite{Fodor:2016pls}. The magenta IRFP of $N_f=12$ is from~\cite{Cheng:2014jba} and the 
		magenta line of our new non-vanishing $N_f=12$ $\beta$-function is also shown in the $\sim 0.1$ range. }
    \vskip -0.1in	
	\label{fig:CompositeBeta}	
\end{figure}

{\em\textit (1) {Light scalar, perhaps dilaton-like?}}
Near-conformal strong dynamics with spontaneous chiral symmetry breaking ($\chi SB$) 
is focused on its emergent light scalar with $0^{++}$ quantum numbers of the $\sigma$-meson, 
perhaps with dilaton-like properties.
With early results reviewed in~\cite{Kuti:2014epa}, this paradigm is very 
different from scaled up Quantum Chromodynamics (QCD) which was the prototype of old Higgs-less Technicolor.
Comparing near-conformal models, with details explained in Figure~\ref{fig:CompositeBeta},
a light composite scalar of the massless SU(2) flavor doublet in the sextet fermion 
representation of SU(3) color was reported in~\cite{Kuti:2014epa,Fodor:2014pqa} whereas the
$N_f=8$ light scalar with fermions in the fundamental representation was 
discovered in~\cite{Aoki:2014oha} and confirmed recently~\cite{Appelquist:2016viq}.
The sextet model $\beta$-function,
with the minimal flavor doublet required for the 
composite Higgs mechanism, indicates the closest position to the lower edge of the conformal window (CW) 
among recently investigated SU(3) gauge theories,
exhibiting the lightest scalar accordingly. The $\beta$-function of the sextet theory with three massless
flavors has a weakly coupled conformal fixed point close to the upper end of the CW~\cite{Ryttov:2010iz}
with apparent crossing into the CW between two and three flavors.
In contrast, uncertainties in crossing into the CW  with fermions in the fundamental representation 
appear to extend into the wider $N_f=8\textendash 12$ flavor range. 
For example, it is not known if for more than eight flavors  the theory gets very close to the CW 
with a much lighter scalar mass than at $N_f=8$.
Based on the findings of~\cite{Cheng:2014jba} and 
a similar zero in the $\beta$-function reported earlier~\cite{Appelquist:2007hu,Appelquist:2009ty},
the $N_f=12$ model has been investigated
as a composite Higgs model built on a conformal fixed point inside the CW~\cite{Brower:2015owo}.
The importance of the question warrants independent determination.

{\em\textit {(2)  PNGB with partial compositeness?}}
Challenges for the near-conformal light scalar paradigm
to generate fermion masses and Yukawa couplings motivates the alternate
PNGB scenario with a massless scalar boson emerging from vacuum misalignment 
of $\chi SB$ as reviewed recently~\cite{Ferretti:2016upr}. 
Model studies with a parametrically light Higgs based on $N_f=n_f+\nu_f$ fermion flavors in the fundamental representation 
of the SU(3) color gauge group could address
the hierarchy problem and fermion mass generation with 
partial compositeness, if $N_f$ is large enough to bring the theory inside
the CW before mass deformations of conformal symmetries are turned on~\cite{Ferretti:2016upr,Vecchi:2015fma,Ma:2015gra}.
For the simple choice $n_f=4$,
the global flavor symmetry SU(4)$\times$SU(4)
is broken to the diagonal SU(4) flavor group and a Higgs-like scalar state is identified
in the PNGB set via $\chi SB$. The custodial SO(4) symmetry of the Standard Model 
remains protected~\cite{Vecchi:2015fma,Ma:2015gra} while a large enough $\nu_f$ is required
to bring the theory close to a strongly coupled IRFP  with expectations of
large baryon anomalous dimensions as the key ingredients of partial compositeness.
The $N_f=12$ choice with $n_f=4$ and $\nu_f=8$  for this PNGB paradigm is discussed in~\cite{Brower:2015owo}
building on the conformal fixed point of twelve flavors, warranting again independent confirmation.

{\bf\begin{center} {II. LATTICE IMPLEMENTATION OF THE STEP $\boldsymbol{\beta}$-FUNCTION}\end{center}}
%
 The gradient flow based diffusion of the gauge fields of  lattice configurations from 
 Hybrid Monte Carlo (HMC) simulations became the method of choice  for studying renormalization 
 effects with great accuracy
  ~\cite{Morningstar:2003gk, Narayanan:2006rf, Luscher:2009eq, Luscher:2010iy, Luscher:2010we,Luscher:2011bx, Lohmayer:2011si}.
In particular, we adapted the method and introduced the 
scale-dependent renormalized gauge coupling $g^2(L)$ where the scale is 
set by the linear size $L$ of the finite volume~\cite{Fodor:2012td,Fodor:2012qh}. This implementation is based on
the gauge invariant trace of the non-Abelian quadratic field strength,
$E(t) = -\frac{1}{2} {\rm Tr} F_{\mu\nu} F_{\mu\nu}(t)$,
renormalized as a composite operator at gradient flow time $t$ on the gauge configurations
and measured from the discretized lattice implementation, as in~\cite{Luscher:2010iy} .
Following~\cite{Fodor:2012td,Fodor:2012qh},  we define
the one-parameter family of  renormalized non-perturbative gauge couplings 
for strongly coupled gauge theories built on the SU(N) color group 
with $N_f$ massless dynamical fermions,
\bea
\label{g2}
g_c^2(t(L)) = \frac{128\pi^2\langle t^2 E(t) \rangle}{3(N^2-1)(1+\delta(c))} ~,
\eea
where the volume-dependent gradient flow time $t(L)$ is set by 
the constant $c = \sqrt{8t}/L$ from the one-parameter family of renormalization schemes,
with $c=0.2$ chosen in this work.
The  factor
\bea
\label{delta}
\delta(c) = - \frac{c^4 \pi^2}{3} + \vartheta^4\left(e^{-1/c^2}\right) - 1
\eea
in Eq.~(\ref{g2}) is chosen to match $g_c^2(t(L))$ to the conventional coupling 
$g_{\msbar} ^2(t(L))$
 in  leading order of perturbation theory for any choice of $c$ and with periodic boundary conditions 
 for the gauge fields in all four directions. 
The origin of the 3rd Jacobi elliptic function $\vartheta$ in Eq.~(\ref{delta}) was explained in~\cite{Fodor:2012td}
including the treatment of zero modes from periodic gauge fields in finite volumes ~\cite{Luscher:1982ma,vanBaal:1985cq,vanBaal:1988va,Coste:1985mn,Coste:1986cb}. 

A scale-dependent renormalized gauge coupling $g^2(L)$  was introduced earlier to probe the
step $\beta$-function, defined as $( g^2(sL) - g^2(L) ) / \log( s^2 )$ for some preset finite scale change $s$
in the linear physical size $L$
of the  four-dimensional volume in the continuum limit of 
lattice discretization~\cite{Luscher:1991wu, Luscher:1992an}. 
The gauge coupling $g^2(L)$ for the determination of the step $\beta$-function is identified in our case with
the definition in ~Eq.~(\ref{g2}) as we drop the preset label $c$ in the notation and $t(L)$ is simply replaced by
$L$.
The renormalization scheme with the preset choice $c=0.2$ and the preset scale factor $s=2$ in our work is identical 
to the one of the previous study~\cite{Cheng:2014jba} including the boundary conditions on gauge fields and fermion fields.
In the continuum limit, the monotonic function $g^2(L)$ implies in any of the volume-dependent schemes that a selected
value of the renormalized gauge coupling sets the physical size $L$ measured in some particular dimensionful physical unit.
Fixed physical size $L$ on the lattice is equivalent to holding $g^2(L)$ fixed at some selected value as
the lattice spacing $a$ is varied and the fixed physical length $L$ is held by the variation of the 
dimensionless linear scale $L/a$ as the bare lattice coupling is tuned without changing the selected fixed value of the
renormalized gauge coupling.
The continuum limit at fixed $g^2(L)$ is obtained by $a^2/L^2\to 0$ extrapolation 
of the residual cut-off dependence in the step $\beta$-function at the target gauge coupling.

In the convention we use, asymptotic freedom in the UV regime corresponds to a positive step $\beta$-function given by
the perturbative loop expansion for small values of the renormalized coupling.
In the infinitesimal derivative limit $s\!\to\!1$ the step  $\beta$-function turns into the conventional one.
If the conventional  $\beta$-function of the theory possesses a fixed point, the step $\beta$-function will
have a zero at the same critical gauge coupling $g_*^2$ as well. 
The scale-dependence of the gauge coupling $g^2(L)$ can be determined from repeated application of the step $\beta$-function starting at some scale $L_0$ set by the initial gauge coupling $g^2(L_0)$ we choose.

%
 {\bf\begin{center} {III. BSM MODELS CLOSE TO THE CONFORMAL WINDOW}\end{center}}
 
 The effect of near-conformal behavior on the light scalar mass is shown in
 Figure~\ref{fig:CompositeBeta}, if the size of the non-perturbative $\beta$-function 
 is used at strong coupling as an indicator for the approach to the CW in the fundamental and sextet
 representations of massless fermions. 
 The mass of the light $\sigma$-like $0^{++}$ scalar particle, as 
 a composite Higgs candidate when coupled to the Electroweak sector, is displayed in units 
 of the Goldstone decay constant $F$ in the massless fermion limit of $\chi SB$ as determined from 
 spectroscopy in each model. The striking trend of decreasing scalar mass is well established 
 as the CW is approached.
 In BSM applications $F=250~GeV$ sets the scale in physical units~\cite{Kuti:2014epa}. 
 The sextet model has the smallest non-zero $\beta$-function 
 relative to the other theories in the fundamental representation, together with the lightest scalar.
 The possibility of the $N_f=12$ model being even closer to the CW with an even lighter 
 scalar is open, if the model is near-conformal without IRFP. 
 Our goal is an independent determination of the fate of the $N_f=12$ IRFP reported earlier~\cite{Cheng:2014jba}.

\newpage

{\bf\begin{center} {IV. $\boldsymbol{N_f = 12}$ SIMULATIONS \\ WITH TARGETED RUN SETS} \end{center}}
 
 The algorithmic details of our new $N_f=12$ simulations are similar to~\cite{Fodor:2012td, Fodor:2015baa}.
 Periodic boundary conditions already defined on the gauge fields, the fermion fields are chosen to be
 anti-periodic in all four directions.
 We utilize the staggered fermion action with massless fermions
 and 4 steps of stout smearing
 with stout parameter $\varrho = 0.12$ on the gauge links~\cite{Morningstar:2003gk}. 
 The gauge action is the tree-level improved 
 Symanzik action \cite{Symanzik:1983dc, Luscher:1984xn}.
 The evolution along a trajectory of the Hybrid Monte Carlo algorithm \cite{Duane:1987de} is implemented
 with multiple time scales~\cite{Sexton:1992nu} and Omelyan integrator~\cite{Takaishi:2005tz}.
 For integration along the gradient flow we use the tree-level improved Symanzik action based
 discretization scheme. The observable $E(t)$ is discretized as in~\cite{Luscher:2010iy}. 
 
 The final 28 runs of Table~\ref{table:data}  ranged in length between 5,000 and 20,000 time units of molecular dynamics.
 The statistical analysis of the renormalized gauge coupling of each run followed~\cite{Wolff:2003sm} 
 and used similar software. Autocorrelation times were measured
 for each run in two independent ways, using estimates from the autocorrelation function of each run,
 and from Jackknifed blocking procedure. Errors on the renormalized couplings were
 consistent from the two procedures and the one from autocorrelation functions is listed in Table~\ref{table:data}.
 Each run went through thermalization and these segments were not included in the analysis.
 For detection of residual thermalization effects the replica method of~\cite{Wolff:2003sm}  
 was used in the analysis. All 28 runs passed Q value tests when mean values and statistical errors of the replica segments 
 were compared for thermal and other variations. 
 \begin{table}[htb] 
 	\begin{center} 
 		\begin{tabular}{|c|c|c||c|c||c|c|}
 			\hline 
 			&\multicolumn{2}{c||}{Target A}&\multicolumn{2}{c||}{Target B}& \multicolumn{2}{c|}{Target C}\\			
 			\hline\hline
 			L/a& $6/g_0^2$& $g^2$ & $6/g_0^2$ & $g^2$ & $6/g_0^2$ & $g^2$  \\			
 			\hline		
 			16 & 3.1519 & 5.9801(29)  & 3.0830&6.1786(39)  & 3.0110 & 6.3930(30)  \\
 			\hline 
 			32 & 3.1519 & 5.9952(79)  &3.0830 & 6.1597(64)  & 3.0110 & 6.3233(74) \\
 			\hline \hline
 			18 & 3.1510 & 5.9767(40)  & 3.0785  & 6.1871(37) & 3.0055  & 6.3909(51)    \\
 			\hline 
 			36 & 3.1510 & 6.0101(71)  & 3.0785 & 6.1840(81) & 3.0055  & 6.3446(64)  \\	
 			\hline\hline
 			20 & 3.1499 & 5.9828(64)  & 3.0704 & 6.1922(64) & 2.9896 & 6.3942(59)   \\
 			\hline 
 			40 & 3.1499  & 6.0419(73) &3.0704 & 6.2137(67)  & 2.9896 & 6.4000(67)  \\
 			\hline \hline
 			24 & 3.1480 & 5.9784(68)  & 3.0680 & 6.1861(55) & 2.9800  & 6.3976(60)  \\
 			\hline 
 			48 & 3.1480 & 6.0758(84)  & 3.0680  & 6.2497(109)  & 2.9800 & 6.4404(122)  \\	
 			\hline\hline		
 			28 &  &   & 3.0698& 6.1839(58)  & 2.9819  & 6.3900(37)  \\
 			\hline 
 			56 &  &   &3.0698 & 6.2792(142) & 2.9819  & 6.4610(124)  \\	
 			\hline\hline					
 			
 		\end{tabular} 
 		\caption{The final 28 runs are tabulated with 14 tuned runs and 14 paired steps.
 			\label{table:data} 
 		} 
 	\end{center} 
 \end{table}
\begin{figure}[htb!]
	\includegraphics[width=1\linewidth]{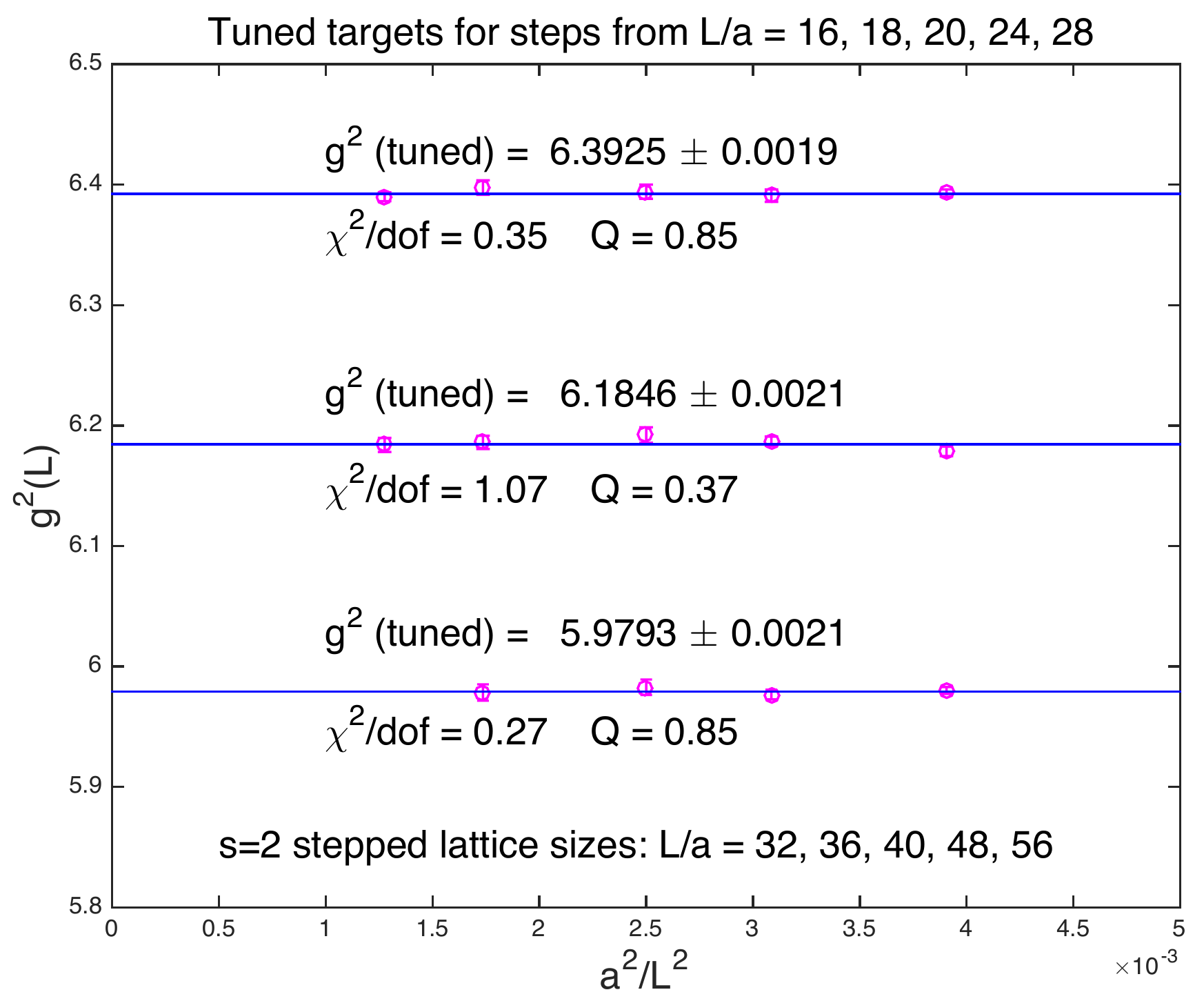}
	\caption{\label{fig:tuning} The statistical significance of precise tuning to three targeted gauge couplings is shown
	  by fitting a constant to each $g^2$ at the lower $L/a$ values of the steps.}
	\vskip -0.2in
\end{figure}

 We targeted the step $\beta$-function at three preselected values of the renormalized gauge coupling to cover the 
 interval where the IRFP was reported \cite{Cheng:2014jba}. In Table~\ref{table:data} results are shown for gauge ensembles 
 from the three target groups A, B, C of the final run sets.
 The 28 runs were grouped into 14 steps of pairs where the lower $L/a$ value was 
  precisely tuned to the target value of the renormalized gauge coupling. The higher $L/a$ volume at the doubled physical
  size determined the step $\beta$-function at finite lattice spacing.
 The first group with 4 steps is target A at $g^2(L)=5.979(2)$ 
 with $L/a=16\!\rightarrow\!32, 18\!\rightarrow\!36, 20\!\rightarrow\!40, 24\!\rightarrow\!48$.
 Both target B at $g^2(L)=6.185(2)$ and target C  at $g^2(L)=6.393(2)$ have 
 an added  fifth step of $L/a=28\!\rightarrow\!56$ for more robust continuum extrapolation. 
 Precise tuning for  $g_0^2$ of the 14 steps of the three targets eliminated
 the largest systematic  uncertainty in the step $\beta$-function from model-dependent interpolation in the bare gauge coupling. 
 Figure~\ref{fig:tuning} shows the remarkable accuracy of tuning for the three targets at better 
 than per mille  accuracy level, like for the entries of Table~\ref{table:data}. 

{\bf\begin{center} {V. CONTINUUM EXTRAPOLATION \\ OF THE STEP $\boldsymbol{\beta}$-FUNCTION}\end{center}}

Cut-off effects have to be removed from the step $\beta$-functions at finite lattice spacing. 
\begin{figure}[htb!]
	\includegraphics[width=0.8\linewidth]{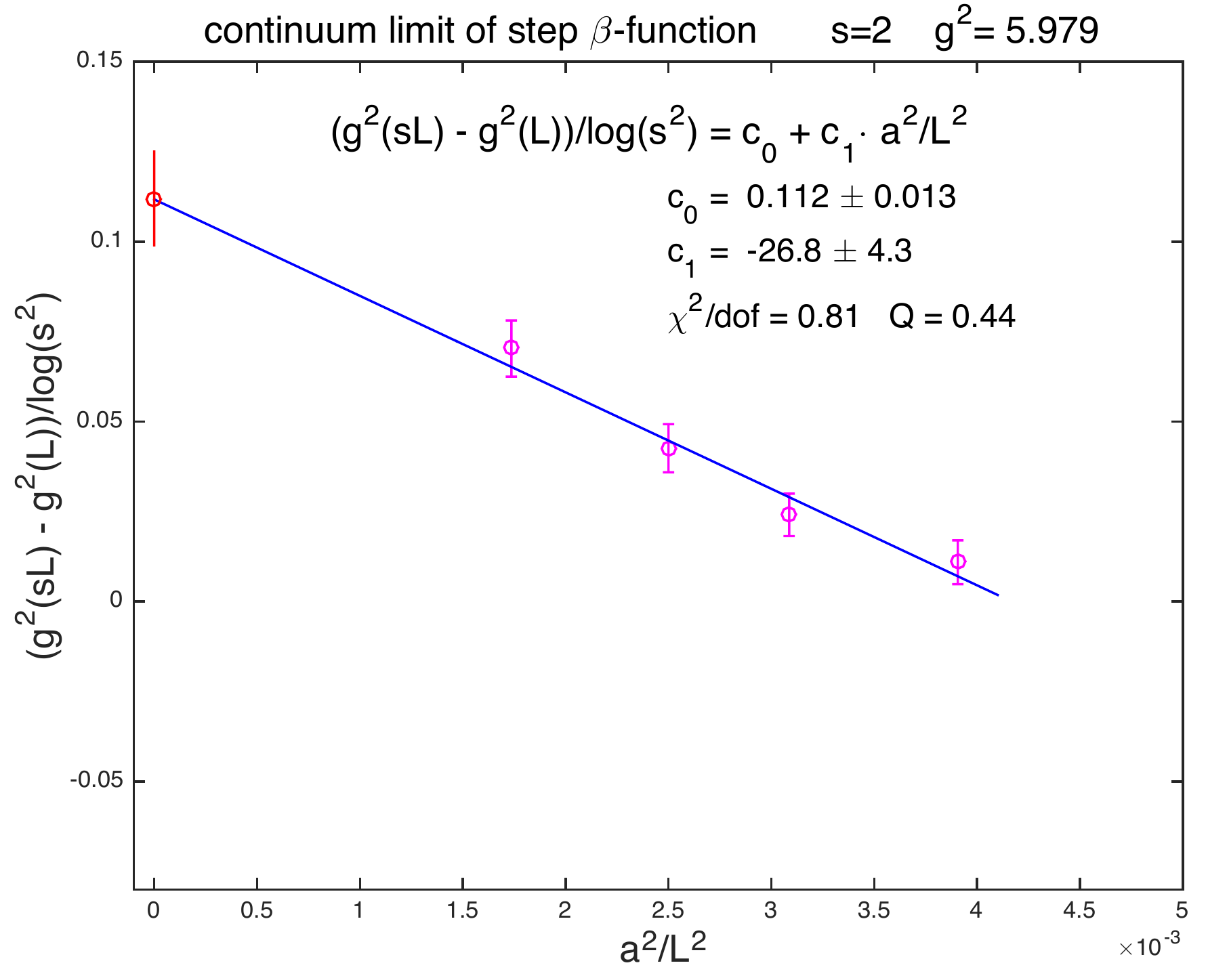}\\
	\includegraphics[width=0.8\linewidth]{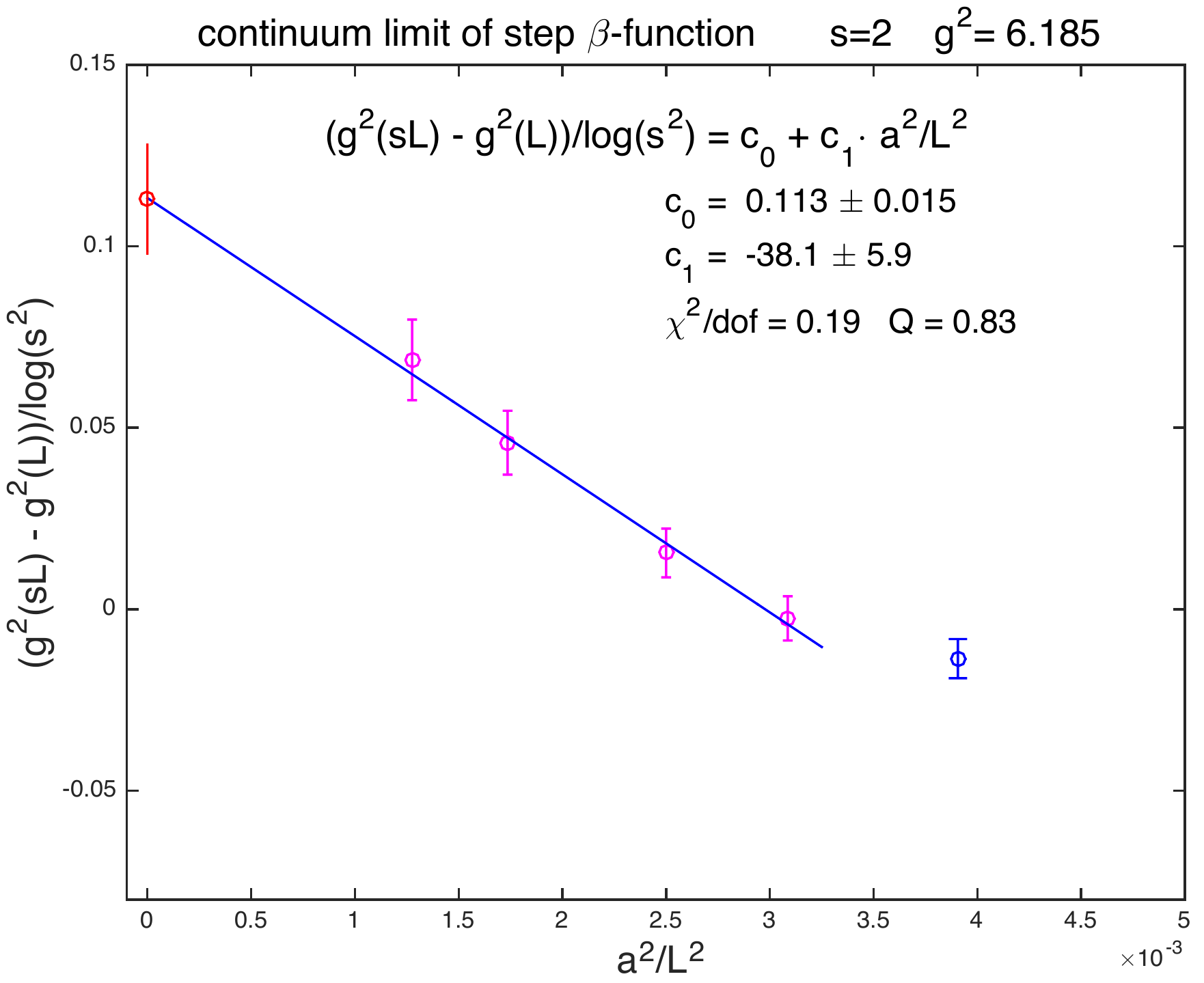}\\
	\includegraphics[width=0.8\linewidth]{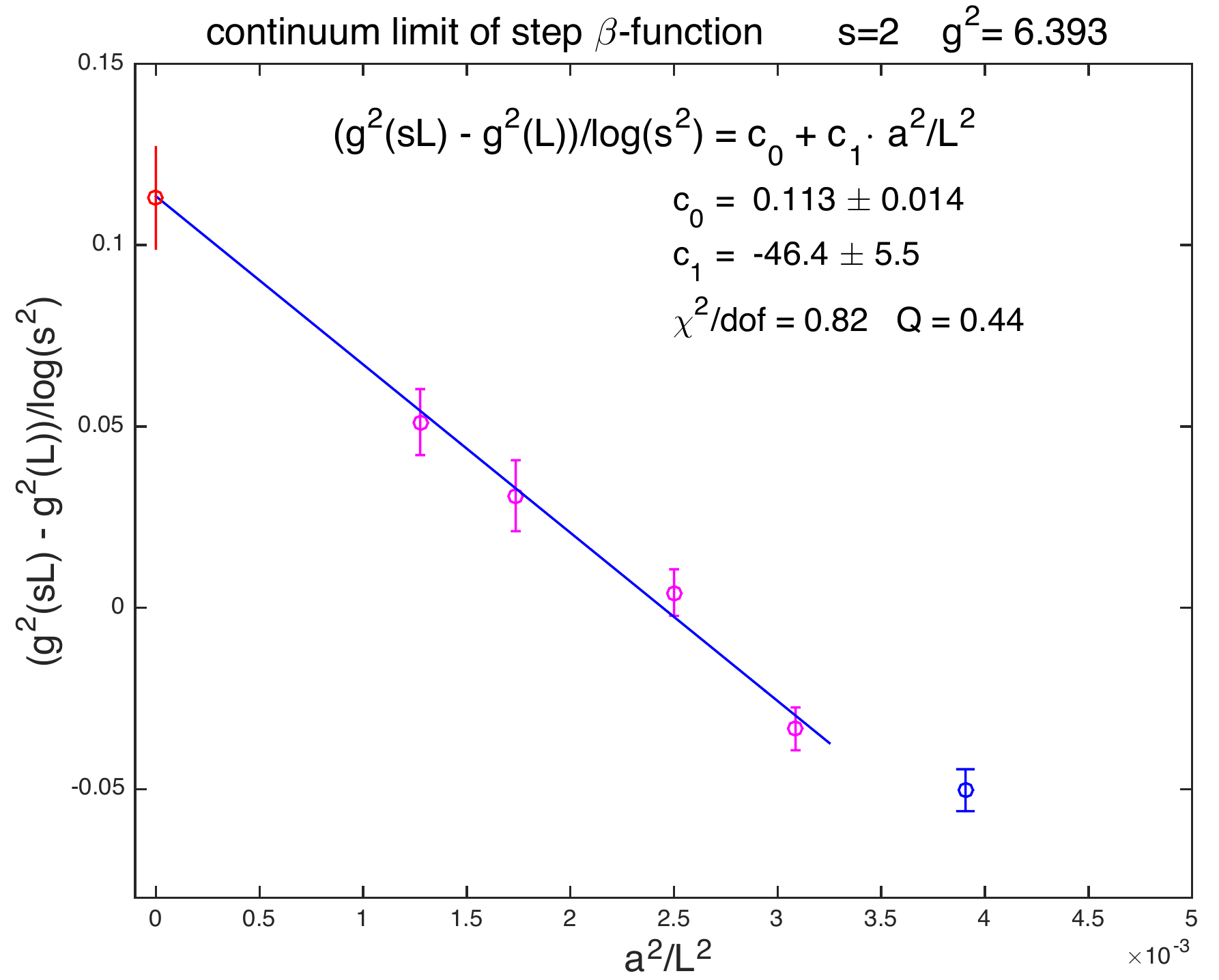}\\	
	\caption{\label{fig:Tuning1} Linear fits in $a^2/L^2$ are shown as explained in the text. The $16\to 32$ steps of target B
		and target C  are not included in the 4-point fits without any influence on the
		overwhelming statistical significance of the results.
		When they are included, 
		the continuum step $\beta$-function
	    drops lower by approximately one standard deviation with comparable errors and increased $\rm {\chi2/dof \sim 1.5}$,  
	   perhaps hinting at sub-leading small $a^4/L^4$
	   cutoff corrections at low $L/a$ when the renormalized gauge coupling gets stronger. }
	\vskip -0.18in
\end{figure}
The leading
cut-off effects are $a^2/L^2$ corrections in each  $L/a\rightarrow 2L/a$ pair for the step $\beta$-function 
at the targeted renormalized couplings.  Linear fits to the lattice step functions in $a^2/L^2$ allows 
continuum extrapolation to the  $a^2/L^2 \rightarrow 0$ limit, as shown in Figure~\ref{fig:Tuning1}.
For all three targets linear four-point fits of the step functions were used with consistently good $\chi^2$ results. 
\begin{figure}[htb!]
\includegraphics[width=3.2in]{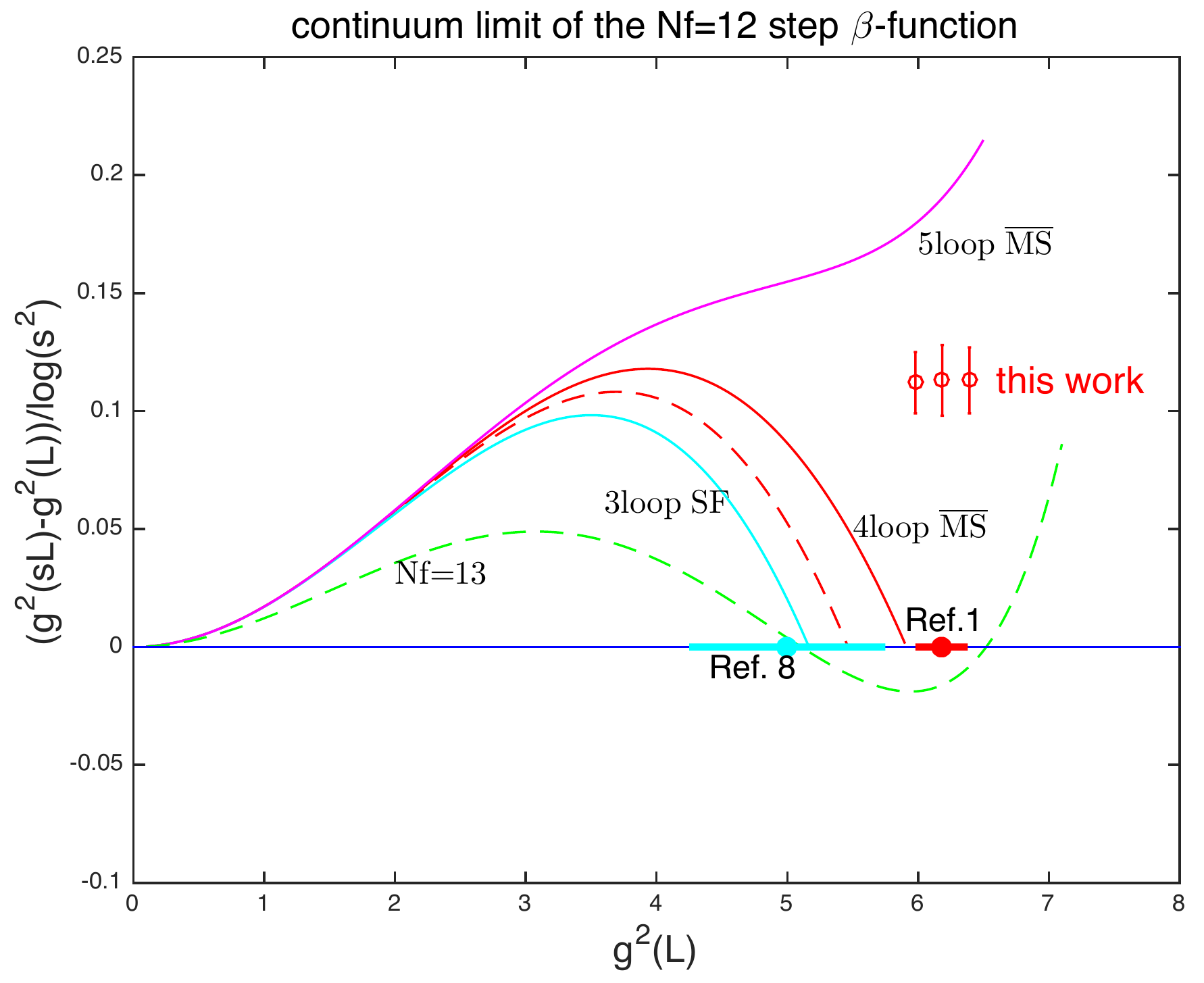}
\caption{\label{fig:Tuning2} The conformal fixed point of~\cite{Cheng:2014jba} 
and the three data points of  our step $\beta$-function are shown (red color).
The IRFP from~\cite{Appelquist:2009ty} (cyan color) and the new 
5-loop $\overline{MS}$ step $\beta$-function of thirteen flavors (dashed green) are discussed in the text.}
%
\vskip -0.15in
\end{figure}
The final results of our continuum step $\beta$-function are shown in Figure~\ref{fig:Tuning2}
with overwhelming statistical evidence against the IRFP of~\cite{Cheng:2014jba} in the targeted
interval.
Leaving open the existence of the IRFP in~\cite{Cheng:2014jba},
a new study of the $\beta$-function appeared recently in a different renormalization scheme of the model
and without our targeted goal~\cite{Lin:2015zpa}.

%
{\bf\begin{center} {VI. NEW DEVELOPMENTS \\ AND CONCLUSIONS} \end{center}}

Originally the zero of the $\beta$-function for twelve flavors was reported at a somewhat
lower value of $g^2$ using the Schr\"odinger functional (SF)
based scheme in agreement with its 3-loop step $\beta$-function~\cite{Appelquist:2009ty}, 
as shown in Figure~\ref{fig:Tuning2} (cyan color). In comparison, the dashed red line 
is the 3-loop prediction of the $\overline{MS}$ scheme within the simulation error of the IRFP.
The 4-loop $\overline{MS}$ result only slightly shifts the prediction and is closer to~\cite{Cheng:2014jba}.
%
%
Although in two different schemes, tantalizing agreement of the simulations and the 
loop expansion lead to the widely held view that twelve massless fermion flavors 
in QCD bring the theory inside the CW.


In a significant new development, the first $\overline{MS}$ calculation of the 5-loop $\beta$-function 
was completed for arbitrary flavor number in QCD~\cite{Baikov:2016tgj}.
Based on the new 5-loop results, it was immediately recognized that
the zero in the $\beta$-function turns complex and the IRFP disappears
for twelve flavors~\cite{Ryttov:2016ner}, consistent with the plot in Figure~\ref{fig:Tuning2}.
It was also shown that  two fixed points appear in the $\beta$-function for thirteen flavors
like in the intriguing scenario 
of~\cite{Kaplan:2009kr}, with
shifting estimates for the lower edge of the CW and for the flavor dependence of the 
mass anomalous dimension~\cite{Ryttov:2016ner}. Five loop $\overline{MS}$  predicts 
two real zeros at $g^2 = 5.11$ and $g^2 = 6.52$ for thirteen flavors, as shown in Figure~\ref{fig:Tuning2}.            
It did not escape our attention that new lattice studies 
of the running coupling with thirteen flavors would be within easy reach 
of the 5-loop $\overline{MS}$ predictions.

Credible proof of conformal behavior based on the $\beta$-function requires two necessary steps 
in strongly coupled gauge theories.
First, the critical gauge coupling $g_*^2$ has to be determined where the scheme-dependent $\beta$-function 
vanishes and signals the location of the conformal IRFP. The slope of the $\beta$-function at 
the fixed point is a scheme-independent 
scaling exponent $\omega$ which controls the leading conformal scaling corrections to fermion mass deformations 
close to the IRFP~\cite{DelDebbio:2010hx,DelDebbio:2010ze,Kuti:2014epa,Fodor:2012et}. 
The choice in scheme dependence can move the position of the 
conformal IRFP but cannot destroy its existence, or change the universal scaling exponent $\omega$.
These are very demanding criteria, unmatched in lattice simulations while reporting zeros 
in the $\beta$-function. 
%

{\begin{center}\bf{ACKNOWLEDGMENTS}\end{center}} 
		
	\noindent
	We acknowledge support by the DOE under grant DE-SC0009919,
	by the NSF under grants 0970137 and 1318220, 
	by OTKA under the grant OTKA-NF-104034, and by the Deutsche
	Forschungsgemeinschaft grant SFB-TR 55. Computational resources were provided 
	by USQCD at Fermilab, 
	by the University of Wuppertal, by Juelich Supercomputing Center on Juqueen
	and by the Institute for Theoretical Physics, Eotvos University. 
	We are grateful to Szabolcs Borsanyi for his code development for the BG/Q platform. We are also 
	grateful to Sandor Katz and Kalman Szabo for their CUDA code development.
\vskip -0.15in
		
\bibliography{v2jkNf12}

\begin{thebibliography}{44}%
\makeatletter
\providecommand \@ifxundefined [1]{%
 \@ifx{#1\undefined}
}%
\providecommand \@ifnum [1]{%
 \ifnum #1\expandafter \@firstoftwo
 \else \expandafter \@secondoftwo
 \fi
}%
\providecommand \@ifx [1]{%
 \ifx #1\expandafter \@firstoftwo
 \else \expandafter \@secondoftwo
 \fi
}%
\providecommand \natexlab [1]{#1}%
\providecommand \enquote  [1]{``#1''}%
\providecommand \bibnamefont  [1]{#1}%
\providecommand \bibfnamefont [1]{#1}%
\providecommand \citenamefont [1]{#1}%
\providecommand \href@noop [0]{\@secondoftwo}%
\providecommand \href [0]{\begingroup \@sanitize@url \@href}%
\providecommand \@href[1]{\@@startlink{#1}\@@href}%
\providecommand \@@href[1]{\endgroup#1\@@endlink}%
\providecommand \@sanitize@url [0]{\catcode `\\12\catcode `\$12\catcode
  `\&12\catcode `\#12\catcode `\^12\catcode `\_12\catcode `\%12\relax}%
\providecommand \@@startlink[1]{}%
\providecommand \@@endlink[0]{}%
\providecommand \url  [0]{\begingroup\@sanitize@url \@url }%
\providecommand \@url [1]{\endgroup\@href {#1}{\urlprefix }}%
\providecommand \urlprefix  [0]{URL }%
\providecommand \Eprint [0]{\href }%
\providecommand \doibase [0]{http://dx.doi.org/}%
\providecommand \selectlanguage [0]{\@gobble}%
\providecommand \bibinfo  [0]{\@secondoftwo}%
\providecommand \bibfield  [0]{\@secondoftwo}%
\providecommand \translation [1]{[#1]}%
\providecommand \BibitemOpen [0]{}%
\providecommand \bibitemStop [0]{}%
\providecommand \bibitemNoStop [0]{.\EOS\space}%
\providecommand \EOS [0]{\spacefactor3000\relax}%
\providecommand \BibitemShut  [1]{\csname bibitem#1\endcsname}%
\let\auto@bib@innerbib\@empty
\bibitem [{\citenamefont {Cheng}\ \emph {et~al.}(2014)\citenamefont {Cheng},
  \citenamefont {Hasenfratz}, \citenamefont {Liu}, \citenamefont
  {Petropoulos},\ and\ \citenamefont {Schaich}}]{Cheng:2014jba}%
  \BibitemOpen
  \bibfield  {author} {\bibinfo {author} {\bibfnamefont {A.}~\bibnamefont
  {Cheng}}, \bibinfo {author} {\bibfnamefont {A.}~\bibnamefont {Hasenfratz}},
  \bibinfo {author} {\bibfnamefont {Y.}~\bibnamefont {Liu}}, \bibinfo {author}
  {\bibfnamefont {G.}~\bibnamefont {Petropoulos}}, \ and\ \bibinfo {author}
  {\bibfnamefont {D.}~\bibnamefont {Schaich}},\ }\href {\doibase
  10.1007/JHEP05(2014)137} {\bibfield  {journal} {\bibinfo  {journal} {JHEP}\
  }\textbf {\bibinfo {volume} {05}},\ \bibinfo {pages} {137} (\bibinfo {year}
  {2014})},\ \Eprint {http://arxiv.org/abs/1404.0984} {arXiv:1404.0984
  [hep-lat]} \BibitemShut {NoStop}%
\bibitem [{\citenamefont {Kuti}(2014)}]{Kuti:2014epa}%
  \BibitemOpen
  \bibfield  {author} {\bibinfo {author} {\bibfnamefont {J.}~\bibnamefont
  {Kuti}},\ }\bibfield  {booktitle} {\emph {\bibinfo {booktitle} {{Proceedings,
  31st International Symposium on Lattice Field Theory (Lattice 2013)}}},\
  }\href@noop {} {\bibfield  {journal} {\bibinfo  {journal} {PoS}\ }\textbf
  {\bibinfo {volume} {LATTICE2013}},\ \bibinfo {pages} {004} (\bibinfo {year}
  {2014})}\BibitemShut {NoStop}%
\bibitem [{\citenamefont {Fodor}\ \emph {et~al.}(2014)\citenamefont {Fodor},
  \citenamefont {Holland}, \citenamefont {Kuti}, \citenamefont {Nogradi},\ and\
  \citenamefont {Wong}}]{Fodor:2014pqa}%
  \BibitemOpen
  \bibfield  {author} {\bibinfo {author} {\bibfnamefont {Z.}~\bibnamefont
  {Fodor}}, \bibinfo {author} {\bibfnamefont {K.}~\bibnamefont {Holland}},
  \bibinfo {author} {\bibfnamefont {J.}~\bibnamefont {Kuti}}, \bibinfo {author}
  {\bibfnamefont {D.}~\bibnamefont {Nogradi}}, \ and\ \bibinfo {author}
  {\bibfnamefont {C.~H.}\ \bibnamefont {Wong}},\ }\bibfield  {booktitle} {\emph
  {\bibinfo {booktitle} {{Proceedings, 31st International Symposium on Lattice
  Field Theory (Lattice 2013)}}},\ }\href@noop {} {\bibfield  {journal}
  {\bibinfo  {journal} {PoS}\ }\textbf {\bibinfo {volume} {LATTICE2013}},\
  \bibinfo {pages} {062} (\bibinfo {year} {2014})}\BibitemShut {NoStop}%
\bibitem [{\citenamefont {Aoki}\ \emph {et~al.}(2014)\citenamefont {Aoki} \emph
  {et~al.}}]{Aoki:2014oha}%
  \BibitemOpen
  \bibfield  {author} {\bibinfo {author} {\bibfnamefont {Y.}~\bibnamefont
  {Aoki}} \emph {et~al.} (\bibinfo {collaboration} {LatKMI}),\ }\href {\doibase
  10.1103/PhysRevD.89.111502} {\bibfield  {journal} {\bibinfo  {journal} {Phys.
  Rev.}\ }\textbf {\bibinfo {volume} {D89}},\ \bibinfo {pages} {111502}
  (\bibinfo {year} {2014})}\BibitemShut {NoStop}%
\bibitem [{\citenamefont {Appelquist}\ \emph {et~al.}(2016)\citenamefont
  {Appelquist} \emph {et~al.}}]{Appelquist:2016viq}%
  \BibitemOpen
  \bibfield  {author} {\bibinfo {author} {\bibfnamefont {T.}~\bibnamefont
  {Appelquist}} \emph {et~al.},\ }\href@noop {} {\  (\bibinfo {year} {2016})},\
  \Eprint {http://arxiv.org/abs/1601.04027} {arXiv:1601.04027 [hep-lat]}
  \BibitemShut {NoStop}%
\bibitem [{\citenamefont {Ryttov}\ and\ \citenamefont
  {Shrock}(2011)}]{Ryttov:2010iz}%
  \BibitemOpen
  \bibfield  {author} {\bibinfo {author} {\bibfnamefont {T.~A.}\ \bibnamefont
  {Ryttov}}\ and\ \bibinfo {author} {\bibfnamefont {R.}~\bibnamefont
  {Shrock}},\ }\href {\doibase 10.1103/PhysRevD.83.056011} {\bibfield
  {journal} {\bibinfo  {journal} {Phys. Rev.}\ }\textbf {\bibinfo {volume}
  {D83}},\ \bibinfo {pages} {056011} (\bibinfo {year} {2011})}\BibitemShut
  {NoStop}%
\bibitem [{\citenamefont {Appelquist}\ \emph {et~al.}(2008)\citenamefont
  {Appelquist}, \citenamefont {Fleming},\ and\ \citenamefont
  {Neil}}]{Appelquist:2007hu}%
  \BibitemOpen
  \bibfield  {author} {\bibinfo {author} {\bibfnamefont {T.}~\bibnamefont
  {Appelquist}}, \bibinfo {author} {\bibfnamefont {G.~T.}\ \bibnamefont
  {Fleming}}, \ and\ \bibinfo {author} {\bibfnamefont {E.~T.}\ \bibnamefont
  {Neil}},\ }\href {\doibase 10.1103/PhysRevLett.100.171607} {\bibfield
  {journal} {\bibinfo  {journal} {Phys. Rev. Lett.}\ }\textbf {\bibinfo
  {volume} {100}},\ \bibinfo {pages} {171607} (\bibinfo {year} {2008})},\
  \bibinfo {note} {[Erratum: Phys. Rev. Lett.102,149902(2009)]},\ \Eprint
  {http://arxiv.org/abs/0712.0609} {arXiv:0712.0609 [hep-ph]} \BibitemShut
  {NoStop}%
\bibitem [{\citenamefont {Appelquist}\ \emph {et~al.}(2009)\citenamefont
  {Appelquist}, \citenamefont {Fleming},\ and\ \citenamefont
  {Neil}}]{Appelquist:2009ty}%
  \BibitemOpen
  \bibfield  {author} {\bibinfo {author} {\bibfnamefont {T.}~\bibnamefont
  {Appelquist}}, \bibinfo {author} {\bibfnamefont {G.~T.}\ \bibnamefont
  {Fleming}}, \ and\ \bibinfo {author} {\bibfnamefont {E.~T.}\ \bibnamefont
  {Neil}},\ }\href {\doibase 10.1103/PhysRevD.79.076010} {\bibfield  {journal}
  {\bibinfo  {journal} {Phys. Rev.}\ }\textbf {\bibinfo {volume} {D79}},\
  \bibinfo {pages} {076010} (\bibinfo {year} {2009})},\ \Eprint
  {http://arxiv.org/abs/0901.3766} {arXiv:0901.3766 [hep-ph]} \BibitemShut
  {NoStop}%
\bibitem [{\citenamefont {Brower}\ \emph {et~al.}(2016)\citenamefont {Brower},
  \citenamefont {Hasenfratz}, \citenamefont {Rebbi}, \citenamefont {Weinberg},\
  and\ \citenamefont {Witzel}}]{Brower:2015owo}%
  \BibitemOpen
  \bibfield  {author} {\bibinfo {author} {\bibfnamefont {R.~C.}\ \bibnamefont
  {Brower}}, \bibinfo {author} {\bibfnamefont {A.}~\bibnamefont {Hasenfratz}},
  \bibinfo {author} {\bibfnamefont {C.}~\bibnamefont {Rebbi}}, \bibinfo
  {author} {\bibfnamefont {E.}~\bibnamefont {Weinberg}}, \ and\ \bibinfo
  {author} {\bibfnamefont {O.}~\bibnamefont {Witzel}},\ }\href {\doibase
  10.1103/PhysRevD.93.075028} {\bibfield  {journal} {\bibinfo  {journal} {Phys.
  Rev.}\ }\textbf {\bibinfo {volume} {D93}},\ \bibinfo {pages} {075028}
  (\bibinfo {year} {2016})}\BibitemShut {NoStop}%
\bibitem [{\citenamefont {Ferretti}(2016)}]{Ferretti:2016upr}%
  \BibitemOpen
  \bibfield  {author} {\bibinfo {author} {\bibfnamefont {G.}~\bibnamefont
  {Ferretti}},\ }\href@noop {} {\  (\bibinfo {year} {2016})},\ \Eprint
  {http://arxiv.org/abs/1604.06467} {arXiv:1604.06467 [hep-ph]} \BibitemShut
  {NoStop}%
\bibitem [{\citenamefont {Vecchi}(2015)}]{Vecchi:2015fma}%
  \BibitemOpen
  \bibfield  {author} {\bibinfo {author} {\bibfnamefont {L.}~\bibnamefont
  {Vecchi}},\ }\href@noop {} {\  (\bibinfo {year} {2015})},\ \Eprint
  {http://arxiv.org/abs/1506.00623} {arXiv:1506.00623 [hep-ph]} \BibitemShut
  {NoStop}%
\bibitem [{\citenamefont {Ma}\ and\ \citenamefont
  {Cacciapaglia}(2016)}]{Ma:2015gra}%
  \BibitemOpen
  \bibfield  {author} {\bibinfo {author} {\bibfnamefont {T.}~\bibnamefont
  {Ma}}\ and\ \bibinfo {author} {\bibfnamefont {G.}~\bibnamefont
  {Cacciapaglia}},\ }\href {\doibase 10.1007/JHEP03(2016)211} {\bibfield
  {journal} {\bibinfo  {journal} {JHEP}\ }\textbf {\bibinfo {volume} {03}},\
  \bibinfo {pages} {211} (\bibinfo {year} {2016})}\BibitemShut {NoStop}%
\bibitem [{\citenamefont {Morningstar}\ and\ \citenamefont
  {Peardon}(2004)}]{Morningstar:2003gk}%
  \BibitemOpen
  \bibfield  {author} {\bibinfo {author} {\bibfnamefont {C.}~\bibnamefont
  {Morningstar}}\ and\ \bibinfo {author} {\bibfnamefont {M.~J.}\ \bibnamefont
  {Peardon}},\ }\href {\doibase 10.1103/PhysRevD.69.054501} {\bibfield
  {journal} {\bibinfo  {journal} {Phys. Rev.}\ }\textbf {\bibinfo {volume}
  {D69}},\ \bibinfo {pages} {054501} (\bibinfo {year} {2004})},\ \Eprint
  {http://arxiv.org/abs/hep-lat/0311018} {arXiv:hep-lat/0311018 [hep-lat]}
  \BibitemShut {NoStop}%
\bibitem [{\citenamefont {Narayanan}\ and\ \citenamefont
  {Neuberger}(2006)}]{Narayanan:2006rf}%
  \BibitemOpen
  \bibfield  {author} {\bibinfo {author} {\bibfnamefont {R.}~\bibnamefont
  {Narayanan}}\ and\ \bibinfo {author} {\bibfnamefont {H.}~\bibnamefont
  {Neuberger}},\ }\href {\doibase 10.1088/1126-6708/2006/03/064} {\bibfield
  {journal} {\bibinfo  {journal} {JHEP}\ }\textbf {\bibinfo {volume} {03}},\
  \bibinfo {pages} {064} (\bibinfo {year} {2006})}\BibitemShut {NoStop}%
\bibitem [{\citenamefont {Luscher}(2010{\natexlab{a}})}]{Luscher:2009eq}%
  \BibitemOpen
  \bibfield  {author} {\bibinfo {author} {\bibfnamefont {M.}~\bibnamefont
  {Luscher}},\ }\href {\doibase 10.1007/s00220-009-0953-7} {\bibfield
  {journal} {\bibinfo  {journal} {Commun. Math. Phys.}\ }\textbf {\bibinfo
  {volume} {293}},\ \bibinfo {pages} {899} (\bibinfo {year}
  {2010}{\natexlab{a}})}\BibitemShut {NoStop}%
\bibitem [{\citenamefont {Lüscher}(2010)}]{Luscher:2010iy}%
  \BibitemOpen
  \bibfield  {author} {\bibinfo {author} {\bibfnamefont {M.}~\bibnamefont
  {Lüscher}},\ }\href {\doibase 10.1007/JHEP08(2010)071,
  10.1007/JHEP03(2014)092} {\bibfield  {journal} {\bibinfo  {journal} {JHEP}\
  }\textbf {\bibinfo {volume} {08}},\ \bibinfo {pages} {071} (\bibinfo {year}
  {2010})}\BibitemShut {NoStop}%
\bibitem [{\citenamefont {Luscher}(2010{\natexlab{b}})}]{Luscher:2010we}%
  \BibitemOpen
  \bibfield  {author} {\bibinfo {author} {\bibfnamefont {M.}~\bibnamefont
  {Luscher}},\ }\bibfield  {booktitle} {\emph {\bibinfo {booktitle}
  {{Proceedings, 28th International Symposium on Lattice field theory (Lattice
  2010)}}},\ }\href@noop {} {\bibfield  {journal} {\bibinfo  {journal} {PoS}\
  }\textbf {\bibinfo {volume} {LATTICE2010}},\ \bibinfo {pages} {015} (\bibinfo
  {year} {2010}{\natexlab{b}})},\ \Eprint {http://arxiv.org/abs/1009.5877}
  {arXiv:1009.5877 [hep-lat]} \BibitemShut {NoStop}%
\bibitem [{\citenamefont {Luscher}\ and\ \citenamefont
  {Weisz}(2011)}]{Luscher:2011bx}%
  \BibitemOpen
  \bibfield  {author} {\bibinfo {author} {\bibfnamefont {M.}~\bibnamefont
  {Luscher}}\ and\ \bibinfo {author} {\bibfnamefont {P.}~\bibnamefont
  {Weisz}},\ }\href {\doibase 10.1007/JHEP02(2011)051} {\bibfield  {journal}
  {\bibinfo  {journal} {JHEP}\ }\textbf {\bibinfo {volume} {02}},\ \bibinfo
  {pages} {051} (\bibinfo {year} {2011})}\BibitemShut {NoStop}%
\bibitem [{\citenamefont {Lohmayer}\ and\ \citenamefont
  {Neuberger}(2011)}]{Lohmayer:2011si}%
  \BibitemOpen
  \bibfield  {author} {\bibinfo {author} {\bibfnamefont {R.}~\bibnamefont
  {Lohmayer}}\ and\ \bibinfo {author} {\bibfnamefont {H.}~\bibnamefont
  {Neuberger}},\ }\bibfield  {booktitle} {\emph {\bibinfo {booktitle}
  {{Proceedings, 29th International Symposium on Lattice field theory (Lattice
  2011)}}},\ }\href@noop {} {\bibfield  {journal} {\bibinfo  {journal} {PoS}\
  }\textbf {\bibinfo {volume} {LATTICE2011}},\ \bibinfo {pages} {249} (\bibinfo
  {year} {2011})},\ \Eprint {http://arxiv.org/abs/1110.3522} {arXiv:1110.3522
  [hep-lat]} \BibitemShut {NoStop}%
\bibitem [{\citenamefont {Fodor}\ \emph
  {et~al.}(2012{\natexlab{a}})\citenamefont {Fodor}, \citenamefont {Holland},
  \citenamefont {Kuti}, \citenamefont {Nogradi},\ and\ \citenamefont
  {Wong}}]{Fodor:2012td}%
  \BibitemOpen
  \bibfield  {author} {\bibinfo {author} {\bibfnamefont {Z.}~\bibnamefont
  {Fodor}}, \bibinfo {author} {\bibfnamefont {K.}~\bibnamefont {Holland}},
  \bibinfo {author} {\bibfnamefont {J.}~\bibnamefont {Kuti}}, \bibinfo {author}
  {\bibfnamefont {D.}~\bibnamefont {Nogradi}}, \ and\ \bibinfo {author}
  {\bibfnamefont {C.~H.}\ \bibnamefont {Wong}},\ }\href {\doibase
  10.1007/JHEP11(2012)007} {\bibfield  {journal} {\bibinfo  {journal} {JHEP}\
  }\textbf {\bibinfo {volume} {11}},\ \bibinfo {pages} {007} (\bibinfo {year}
  {2012}{\natexlab{a}})},\ \Eprint {http://arxiv.org/abs/1208.1051}
  {arXiv:1208.1051 [hep-lat]} \BibitemShut {NoStop}%
\bibitem [{\citenamefont {Fodor}\ \emph
  {et~al.}(2012{\natexlab{b}})\citenamefont {Fodor}, \citenamefont {Holland},
  \citenamefont {Kuti}, \citenamefont {Nogradi},\ and\ \citenamefont
  {Wong}}]{Fodor:2012qh}%
  \BibitemOpen
  \bibfield  {author} {\bibinfo {author} {\bibfnamefont {Z.}~\bibnamefont
  {Fodor}}, \bibinfo {author} {\bibfnamefont {K.}~\bibnamefont {Holland}},
  \bibinfo {author} {\bibfnamefont {J.}~\bibnamefont {Kuti}}, \bibinfo {author}
  {\bibfnamefont {D.}~\bibnamefont {Nogradi}}, \ and\ \bibinfo {author}
  {\bibfnamefont {C.~H.}\ \bibnamefont {Wong}},\ }\bibfield  {booktitle} {\emph
  {\bibinfo {booktitle} {{Proceedings, 30th International Symposium on Lattice
  Field Theory (Lattice 2012)}}},\ }\href@noop {} {\bibfield  {journal}
  {\bibinfo  {journal} {PoS}\ }\textbf {\bibinfo {volume} {LATTICE2012}},\
  \bibinfo {pages} {050} (\bibinfo {year} {2012}{\natexlab{b}})},\ \Eprint
  {http://arxiv.org/abs/1211.3247} {arXiv:1211.3247 [hep-lat]} \BibitemShut
  {NoStop}%
\bibitem [{\citenamefont {Luscher}(1983)}]{Luscher:1982ma}%
  \BibitemOpen
  \bibfield  {author} {\bibinfo {author} {\bibfnamefont {M.}~\bibnamefont
  {Luscher}},\ }\href {\doibase 10.1016/0550-3213(83)90436-4} {\bibfield
  {journal} {\bibinfo  {journal} {Nucl. Phys.}\ }\textbf {\bibinfo {volume}
  {B219}},\ \bibinfo {pages} {233} (\bibinfo {year} {1983})}\BibitemShut
  {NoStop}%
\bibitem [{\citenamefont {van Baal}(1986)}]{vanBaal:1985cq}%
  \BibitemOpen
  \bibfield  {author} {\bibinfo {author} {\bibfnamefont {P.}~\bibnamefont {van
  Baal}},\ }\href {\doibase 10.1016/0550-3213(86)90497-9} {\bibfield  {journal}
  {\bibinfo  {journal} {Nucl. Phys.}\ }\textbf {\bibinfo {volume} {B264}},\
  \bibinfo {pages} {548} (\bibinfo {year} {1986})}\BibitemShut {NoStop}%
\bibitem [{\citenamefont {van Baal}(1988)}]{vanBaal:1988va}%
  \BibitemOpen
  \bibfield  {author} {\bibinfo {author} {\bibfnamefont {P.}~\bibnamefont {van
  Baal}},\ }\href {\doibase 10.1016/0550-3213(88)90323-9,
  10.1016/0550-3213(89)90582-8} {\bibfield  {journal} {\bibinfo  {journal}
  {Nucl. Phys.}\ }\textbf {\bibinfo {volume} {B307}},\ \bibinfo {pages} {274}
  (\bibinfo {year} {1988})},\ \bibinfo {note} {[Erratum: Nucl.
  Phys.B312,752(1989)]}\BibitemShut {NoStop}%
\bibitem [{\citenamefont {Coste}\ \emph {et~al.}(1985)\citenamefont {Coste},
  \citenamefont {Gonzalez-Arroyo}, \citenamefont {Jurkiewicz},\ and\
  \citenamefont {Korthals~Altes}}]{Coste:1985mn}%
  \BibitemOpen
  \bibfield  {author} {\bibinfo {author} {\bibfnamefont {A.}~\bibnamefont
  {Coste}}, \bibinfo {author} {\bibfnamefont {A.}~\bibnamefont
  {Gonzalez-Arroyo}}, \bibinfo {author} {\bibfnamefont {J.}~\bibnamefont
  {Jurkiewicz}}, \ and\ \bibinfo {author} {\bibfnamefont {C.~P.}\ \bibnamefont
  {Korthals~Altes}},\ }\href {\doibase 10.1016/0550-3213(85)90064-1} {\bibfield
   {journal} {\bibinfo  {journal} {Nucl. Phys.}\ }\textbf {\bibinfo {volume}
  {B262}},\ \bibinfo {pages} {67} (\bibinfo {year} {1985})}\BibitemShut
  {NoStop}%
\bibitem [{\citenamefont {Coste}\ \emph {et~al.}(1987)\citenamefont {Coste},
  \citenamefont {Gonzalez-Arroyo}, \citenamefont {Korthals~Altes},
  \citenamefont {Soderberg},\ and\ \citenamefont {Tarancon}}]{Coste:1986cb}%
  \BibitemOpen
  \bibfield  {author} {\bibinfo {author} {\bibfnamefont {A.}~\bibnamefont
  {Coste}}, \bibinfo {author} {\bibfnamefont {A.}~\bibnamefont
  {Gonzalez-Arroyo}}, \bibinfo {author} {\bibfnamefont {C.~P.}\ \bibnamefont
  {Korthals~Altes}}, \bibinfo {author} {\bibfnamefont {B.}~\bibnamefont
  {Soderberg}}, \ and\ \bibinfo {author} {\bibfnamefont {A.}~\bibnamefont
  {Tarancon}},\ }\href {\doibase 10.1016/0550-3213(87)90118-0} {\bibfield
  {journal} {\bibinfo  {journal} {Nucl. Phys.}\ }\textbf {\bibinfo {volume}
  {B287}},\ \bibinfo {pages} {569} (\bibinfo {year} {1987})}\BibitemShut
  {NoStop}%
\bibitem [{\citenamefont {Luscher}\ \emph {et~al.}(1991)\citenamefont
  {Luscher}, \citenamefont {Weisz},\ and\ \citenamefont
  {Wolff}}]{Luscher:1991wu}%
  \BibitemOpen
  \bibfield  {author} {\bibinfo {author} {\bibfnamefont {M.}~\bibnamefont
  {Luscher}}, \bibinfo {author} {\bibfnamefont {P.}~\bibnamefont {Weisz}}, \
  and\ \bibinfo {author} {\bibfnamefont {U.}~\bibnamefont {Wolff}},\ }\href
  {\doibase 10.1016/0550-3213(91)90298-C} {\bibfield  {journal} {\bibinfo
  {journal} {Nucl. Phys.}\ }\textbf {\bibinfo {volume} {B359}},\ \bibinfo
  {pages} {221} (\bibinfo {year} {1991})}\BibitemShut {NoStop}%
\bibitem [{\citenamefont {Luscher}\ \emph {et~al.}(1992)\citenamefont
  {Luscher}, \citenamefont {Narayanan}, \citenamefont {Weisz},\ and\
  \citenamefont {Wolff}}]{Luscher:1992an}%
  \BibitemOpen
  \bibfield  {author} {\bibinfo {author} {\bibfnamefont {M.}~\bibnamefont
  {Luscher}}, \bibinfo {author} {\bibfnamefont {R.}~\bibnamefont {Narayanan}},
  \bibinfo {author} {\bibfnamefont {P.}~\bibnamefont {Weisz}}, \ and\ \bibinfo
  {author} {\bibfnamefont {U.}~\bibnamefont {Wolff}},\ }\href {\doibase
  10.1016/0550-3213(92)90466-O} {\bibfield  {journal} {\bibinfo  {journal}
  {Nucl. Phys.}\ }\textbf {\bibinfo {volume} {B384}},\ \bibinfo {pages} {168}
  (\bibinfo {year} {1992})},\ \Eprint {http://arxiv.org/abs/hep-lat/9207009}
  {arXiv:hep-lat/9207009 [hep-lat]} \BibitemShut {NoStop}%
\bibitem [{\citenamefont {Fodor}\ \emph
  {et~al.}(2015{\natexlab{a}})\citenamefont {Fodor}, \citenamefont {Holland},
  \citenamefont {Kuti}, \citenamefont {Mondal}, \citenamefont {Nogradi},\ and\
  \citenamefont {Wong}}]{Fodor:2015baa}%
  \BibitemOpen
  \bibfield  {author} {\bibinfo {author} {\bibfnamefont {Z.}~\bibnamefont
  {Fodor}}, \bibinfo {author} {\bibfnamefont {K.}~\bibnamefont {Holland}},
  \bibinfo {author} {\bibfnamefont {J.}~\bibnamefont {Kuti}}, \bibinfo {author}
  {\bibfnamefont {S.}~\bibnamefont {Mondal}}, \bibinfo {author} {\bibfnamefont
  {D.}~\bibnamefont {Nogradi}}, \ and\ \bibinfo {author} {\bibfnamefont
  {C.~H.}\ \bibnamefont {Wong}},\ }\href {\doibase 10.1007/JHEP06(2015)019}
  {\bibfield  {journal} {\bibinfo  {journal} {JHEP}\ }\textbf {\bibinfo
  {volume} {06}},\ \bibinfo {pages} {019} (\bibinfo {year}
  {2015}{\natexlab{a}})},\ \Eprint {http://arxiv.org/abs/1503.01132}
  {arXiv:1503.01132 [hep-lat]} \BibitemShut {NoStop}%
\bibitem [{\citenamefont {Fodor}\ \emph
  {et~al.}(2015{\natexlab{b}})\citenamefont {Fodor}, \citenamefont {Holland},
  \citenamefont {Kuti}, \citenamefont {Mondal}, \citenamefont {Nogradi},\ and\
  \citenamefont {Wong}}]{Fodor:2015zna}%
  \BibitemOpen
  \bibfield  {author} {\bibinfo {author} {\bibfnamefont {Z.}~\bibnamefont
  {Fodor}}, \bibinfo {author} {\bibfnamefont {K.}~\bibnamefont {Holland}},
  \bibinfo {author} {\bibfnamefont {J.}~\bibnamefont {Kuti}}, \bibinfo {author}
  {\bibfnamefont {S.}~\bibnamefont {Mondal}}, \bibinfo {author} {\bibfnamefont
  {D.}~\bibnamefont {Nogradi}}, \ and\ \bibinfo {author} {\bibfnamefont
  {C.~H.}\ \bibnamefont {Wong}},\ }\href {\doibase 10.1007/JHEP09(2015)039}
  {\bibfield  {journal} {\bibinfo  {journal} {JHEP}\ }\textbf {\bibinfo
  {volume} {09}},\ \bibinfo {pages} {039} (\bibinfo {year}
  {2015}{\natexlab{b}})},\ \Eprint {http://arxiv.org/abs/1506.06599}
  {arXiv:1506.06599 [hep-lat]} \BibitemShut {NoStop}%
\bibitem [{\citenamefont {Fodor}\ \emph {et~al.}(2016)\citenamefont {Fodor},
  \citenamefont {Holland}, \citenamefont {Kuti}, \citenamefont {Mondal},
  \citenamefont {Nogradi},\ and\ \citenamefont {Wong}}]{Fodor:2016pls}%
  \BibitemOpen
  \bibfield  {author} {\bibinfo {author} {\bibfnamefont {Z.}~\bibnamefont
  {Fodor}}, \bibinfo {author} {\bibfnamefont {K.}~\bibnamefont {Holland}},
  \bibinfo {author} {\bibfnamefont {J.}~\bibnamefont {Kuti}}, \bibinfo {author}
  {\bibfnamefont {S.}~\bibnamefont {Mondal}}, \bibinfo {author} {\bibfnamefont
  {D.}~\bibnamefont {Nogradi}}, \ and\ \bibinfo {author} {\bibfnamefont
  {C.~H.}\ \bibnamefont {Wong}},\ }in\ \href
  {https://inspirehep.net/record/1466133/files/arXiv:1605.08750.pdf} {\emph
  {\bibinfo {booktitle} {{Proceedings, 33rd International Symposium on Lattice
  Field Theory (Lattice 2015)}}}}\ (\bibinfo {year} {2016})\ \Eprint
  {http://arxiv.org/abs/1605.08750} {arXiv:1605.08750 [hep-lat]} \BibitemShut
  {NoStop}%
\bibitem [{\citenamefont {Symanzik}(1983)}]{Symanzik:1983dc}%
  \BibitemOpen
  \bibfield  {author} {\bibinfo {author} {\bibfnamefont {K.}~\bibnamefont
  {Symanzik}},\ }\href {\doibase 10.1016/0550-3213(83)90468-6} {\bibfield
  {journal} {\bibinfo  {journal} {Nucl. Phys.}\ }\textbf {\bibinfo {volume}
  {B226}},\ \bibinfo {pages} {187} (\bibinfo {year} {1983})}\BibitemShut
  {NoStop}%
\bibitem [{\citenamefont {Luscher}\ and\ \citenamefont
  {Weisz}(1985)}]{Luscher:1984xn}%
  \BibitemOpen
  \bibfield  {author} {\bibinfo {author} {\bibfnamefont {M.}~\bibnamefont
  {Luscher}}\ and\ \bibinfo {author} {\bibfnamefont {P.}~\bibnamefont
  {Weisz}},\ }\href {\doibase 10.1007/BF01206178} {\bibfield  {journal}
  {\bibinfo  {journal} {Commun. Math. Phys.}\ }\textbf {\bibinfo {volume}
  {97}},\ \bibinfo {pages} {59} (\bibinfo {year} {1985})},\ \bibinfo {note}
  {[Erratum: Commun. Math. Phys.98,433(1985)]}\BibitemShut {NoStop}%
\bibitem [{\citenamefont {Duane}\ \emph {et~al.}(1987)\citenamefont {Duane},
  \citenamefont {Kennedy}, \citenamefont {Pendleton},\ and\ \citenamefont
  {Roweth}}]{Duane:1987de}%
  \BibitemOpen
  \bibfield  {author} {\bibinfo {author} {\bibfnamefont {S.}~\bibnamefont
  {Duane}}, \bibinfo {author} {\bibfnamefont {A.~D.}\ \bibnamefont {Kennedy}},
  \bibinfo {author} {\bibfnamefont {B.~J.}\ \bibnamefont {Pendleton}}, \ and\
  \bibinfo {author} {\bibfnamefont {D.}~\bibnamefont {Roweth}},\ }\href
  {\doibase 10.1016/0370-2693(87)91197-X} {\bibfield  {journal} {\bibinfo
  {journal} {Phys. Lett.}\ }\textbf {\bibinfo {volume} {B195}},\ \bibinfo
  {pages} {216} (\bibinfo {year} {1987})}\BibitemShut {NoStop}%
\bibitem [{\citenamefont {Sexton}\ and\ \citenamefont
  {Weingarten}(1992)}]{Sexton:1992nu}%
  \BibitemOpen
  \bibfield  {author} {\bibinfo {author} {\bibfnamefont {J.~C.}\ \bibnamefont
  {Sexton}}\ and\ \bibinfo {author} {\bibfnamefont {D.~H.}\ \bibnamefont
  {Weingarten}},\ }\href {\doibase 10.1016/0550-3213(92)90263-B} {\bibfield
  {journal} {\bibinfo  {journal} {Nucl. Phys.}\ }\textbf {\bibinfo {volume}
  {B380}},\ \bibinfo {pages} {665} (\bibinfo {year} {1992})}\BibitemShut
  {NoStop}%
\bibitem [{\citenamefont {Takaishi}\ and\ \citenamefont
  {de~Forcrand}(2006)}]{Takaishi:2005tz}%
  \BibitemOpen
  \bibfield  {author} {\bibinfo {author} {\bibfnamefont {T.}~\bibnamefont
  {Takaishi}}\ and\ \bibinfo {author} {\bibfnamefont {P.}~\bibnamefont
  {de~Forcrand}},\ }\href {\doibase 10.1103/PhysRevE.73.036706} {\bibfield
  {journal} {\bibinfo  {journal} {Phys. Rev.}\ }\textbf {\bibinfo {volume}
  {E73}},\ \bibinfo {pages} {036706} (\bibinfo {year} {2006})},\ \Eprint
  {http://arxiv.org/abs/hep-lat/0505020} {arXiv:hep-lat/0505020 [hep-lat]}
  \BibitemShut {NoStop}%
\bibitem [{\citenamefont {Wolff}(2004)}]{Wolff:2003sm}%
  \BibitemOpen
  \bibfield  {author} {\bibinfo {author} {\bibfnamefont {U.}~\bibnamefont
  {Wolff}} (\bibinfo {collaboration} {ALPHA}),\ }\href {\doibase
  10.1016/S0010-4655(03)00467-3, 10.1016/j.cpc.2006.12.001} {\bibfield
  {journal} {\bibinfo  {journal} {Comput. Phys. Commun.}\ }\textbf {\bibinfo
  {volume} {156}},\ \bibinfo {pages} {143} (\bibinfo {year} {2004})},\ \bibinfo
  {note} {[Erratum: Comput. Phys. Commun.176,383(2007)]}\BibitemShut {NoStop}%
\bibitem [{\citenamefont {Lin}\ \emph {et~al.}(2015)\citenamefont {Lin},
  \citenamefont {Ogawa},\ and\ \citenamefont {Ramos}}]{Lin:2015zpa}%
  \BibitemOpen
  \bibfield  {author} {\bibinfo {author} {\bibfnamefont {C.~J.~D.}\
  \bibnamefont {Lin}}, \bibinfo {author} {\bibfnamefont {K.}~\bibnamefont
  {Ogawa}}, \ and\ \bibinfo {author} {\bibfnamefont {A.}~\bibnamefont
  {Ramos}},\ }\href {\doibase 10.1007/JHEP12(2015)103} {\bibfield  {journal}
  {\bibinfo  {journal} {JHEP}\ }\textbf {\bibinfo {volume} {12}},\ \bibinfo
  {pages} {103} (\bibinfo {year} {2015})},\ \Eprint
  {http://arxiv.org/abs/1510.05755} {arXiv:1510.05755 [hep-lat]} \BibitemShut
  {NoStop}%
\bibitem [{\citenamefont {Baikov}\ \emph {et~al.}(2016)\citenamefont {Baikov},
  \citenamefont {Chetyrkin},\ and\ \citenamefont {Kühn}}]{Baikov:2016tgj}%
  \BibitemOpen
  \bibfield  {author} {\bibinfo {author} {\bibfnamefont {P.~A.}\ \bibnamefont
  {Baikov}}, \bibinfo {author} {\bibfnamefont {K.~G.}\ \bibnamefont
  {Chetyrkin}}, \ and\ \bibinfo {author} {\bibfnamefont {J.~H.}\ \bibnamefont
  {Kühn}},\ }\href@noop {} {\  (\bibinfo {year} {2016})},\ \Eprint
  {http://arxiv.org/abs/1606.08659} {arXiv:1606.08659 [hep-ph]} \BibitemShut
  {NoStop}%
\bibitem [{\citenamefont {Ryttov}\ and\ \citenamefont
  {Shrock}(2016)}]{Ryttov:2016ner}%
  \BibitemOpen
  \bibfield  {author} {\bibinfo {author} {\bibfnamefont {T.~A.}\ \bibnamefont
  {Ryttov}}\ and\ \bibinfo {author} {\bibfnamefont {R.}~\bibnamefont
  {Shrock}},\ }\href@noop {} {\  (\bibinfo {year} {2016})},\ \Eprint
  {http://arxiv.org/abs/1607.06866} {arXiv:1607.06866 [hep-th]} \BibitemShut
  {NoStop}%
\bibitem [{\citenamefont {Kaplan}\ \emph {et~al.}(2009)\citenamefont {Kaplan},
  \citenamefont {Lee}, \citenamefont {Son},\ and\ \citenamefont
  {Stephanov}}]{Kaplan:2009kr}%
  \BibitemOpen
  \bibfield  {author} {\bibinfo {author} {\bibfnamefont {D.~B.}\ \bibnamefont
  {Kaplan}}, \bibinfo {author} {\bibfnamefont {J.-W.}\ \bibnamefont {Lee}},
  \bibinfo {author} {\bibfnamefont {D.~T.}\ \bibnamefont {Son}}, \ and\
  \bibinfo {author} {\bibfnamefont {M.~A.}\ \bibnamefont {Stephanov}},\ }\href
  {\doibase 10.1103/PhysRevD.80.125005} {\bibfield  {journal} {\bibinfo
  {journal} {Phys. Rev.}\ }\textbf {\bibinfo {volume} {D80}},\ \bibinfo {pages}
  {125005} (\bibinfo {year} {2009})},\ \Eprint {http://arxiv.org/abs/0905.4752}
  {arXiv:0905.4752 [hep-th]} \BibitemShut {NoStop}%
\bibitem [{\citenamefont {Del~Debbio}\ \emph {et~al.}(2010)\citenamefont
  {Del~Debbio}, \citenamefont {Lucini}, \citenamefont {Patella}, \citenamefont
  {Pica},\ and\ \citenamefont {Rago}}]{DelDebbio:2010hx}%
  \BibitemOpen
  \bibfield  {author} {\bibinfo {author} {\bibfnamefont {L.}~\bibnamefont
  {Del~Debbio}}, \bibinfo {author} {\bibfnamefont {B.}~\bibnamefont {Lucini}},
  \bibinfo {author} {\bibfnamefont {A.}~\bibnamefont {Patella}}, \bibinfo
  {author} {\bibfnamefont {C.}~\bibnamefont {Pica}}, \ and\ \bibinfo {author}
  {\bibfnamefont {A.}~\bibnamefont {Rago}},\ }\href {\doibase
  10.1103/PhysRevD.82.014510} {\bibfield  {journal} {\bibinfo  {journal} {Phys.
  Rev.}\ }\textbf {\bibinfo {volume} {D82}},\ \bibinfo {pages} {014510}
  (\bibinfo {year} {2010})},\ \Eprint {http://arxiv.org/abs/1004.3206}
  {arXiv:1004.3206 [hep-lat]} \BibitemShut {NoStop}%
\bibitem [{\citenamefont {Del~Debbio}\ and\ \citenamefont
  {Zwicky}(2010)}]{DelDebbio:2010ze}%
  \BibitemOpen
  \bibfield  {author} {\bibinfo {author} {\bibfnamefont {L.}~\bibnamefont
  {Del~Debbio}}\ and\ \bibinfo {author} {\bibfnamefont {R.}~\bibnamefont
  {Zwicky}},\ }\href {\doibase 10.1103/PhysRevD.82.014502} {\bibfield
  {journal} {\bibinfo  {journal} {Phys. Rev.}\ }\textbf {\bibinfo {volume}
  {D82}},\ \bibinfo {pages} {014502} (\bibinfo {year} {2010})},\ \Eprint
  {http://arxiv.org/abs/1005.2371} {arXiv:1005.2371 [hep-ph]} \BibitemShut
  {NoStop}%
\bibitem [{\citenamefont {Fodor}\ \emph
  {et~al.}(2012{\natexlab{c}})\citenamefont {Fodor}, \citenamefont {Holland},
  \citenamefont {Kuti}, \citenamefont {Nogradi}, \citenamefont {Schroeder},\
  and\ \citenamefont {Wong}}]{Fodor:2012et}%
  \BibitemOpen
  \bibfield  {author} {\bibinfo {author} {\bibfnamefont {Z.}~\bibnamefont
  {Fodor}}, \bibinfo {author} {\bibfnamefont {K.}~\bibnamefont {Holland}},
  \bibinfo {author} {\bibfnamefont {J.}~\bibnamefont {Kuti}}, \bibinfo {author}
  {\bibfnamefont {D.}~\bibnamefont {Nogradi}}, \bibinfo {author} {\bibfnamefont
  {C.}~\bibnamefont {Schroeder}}, \ and\ \bibinfo {author} {\bibfnamefont
  {C.~H.}\ \bibnamefont {Wong}},\ }\bibfield  {booktitle} {\emph {\bibinfo
  {booktitle} {{Proceedings, 30th International Symposium on Lattice Field
  Theory (Lattice 2012)}}},\ }\href@noop {} {\bibfield  {journal} {\bibinfo
  {journal} {PoS}\ }\textbf {\bibinfo {volume} {LATTICE2012}},\ \bibinfo
  {pages} {279} (\bibinfo {year} {2012}{\natexlab{c}})},\ \Eprint
  {http://arxiv.org/abs/1211.4238} {arXiv:1211.4238 [hep-lat]} \BibitemShut
  {NoStop}%
\end{thebibliography}%
	
\end{document}